\documentclass[%
 reprint,
 amsmath,amssymb,
 aps,
 prl,
 floatfix,
 superscriptaddress
]{revtex4-1}

\usepackage{graphicx}
\usepackage{dcolumn}
\usepackage{bm}
\usepackage{multirow}

\graphicspath{{./Figs/}}

\begin{document}

\title{Effects of Fano resonance on optical chirality of planar plasmonic nanodevices}

\author{Yongsop Hwang}
\affiliation{Nanophotonics Research Centre, Shenzhen University, Shenzhen, 518060, China}
\affiliation{School of Engineering, RMIT University, Melbourne, VIC 3001, Australia}
\author{Seojoo Lee}
\affiliation{Department of Physics, Korea University, Seoul 02841, Republic of Korea}
\author{Sejeong Kim}
\affiliation{School of Mathematical and Physical Sciences, University of Technology Sydney, Ultimo, NSW 2007, Australia}
\author{Jiao Lin}
\email{jiao.lin@rmit.edu.au}
\affiliation{Nanophotonics Research Centre, Shenzhen University, Shenzhen, 518060, China}
\affiliation{School of Engineering, RMIT University, Melbourne, VIC 3001, Australia}
\author{Xiao-Cong Yuan}
\email{xcyuan@szu.edu.cn}
\affiliation{Nanophotonics Research Centre, Shenzhen University, Shenzhen, 518060, China}

\date{\today}

\begin{abstract}
The effects of Fano resonance on the optical chirality of planar plasmonic nanodevices in the visible wavelength range are experimentally observed and theoretically explained.
The nanodevice consists of a nanodisk at the center with six gold nanorods with an orientation angle to exhibit optical chirality under dark-field illumination.
The chiral response induced by the gold nanorods are affected by the presence of the nanodisk with different diameters which causes Fano resonance.
An intriguing change to the opposite selection preference of different handedness of the circularly polarized light has been clearly observed experimentally.
This change of the preference is understood based on the extended coupled oscillator model.
Moreover, electrostatic analysis and the time-dependent simulations provide a further understanding of the phenomenon.
The observed and understood effects of Fano resonance on optical chirality enables effective manipulation of chiral characteristics of planar subwavelength nanodevices.
\end{abstract}

\maketitle

Optical chirality which exhibits different responses between right- and left-handed circularly-polarized (RCP and LCP) light has attracted attention due to its intriguing physical characteristics~\cite{pendry2004chiral} as well as its applications, for instance, in pharmaceuticals~\cite{nguyen2006chiral} and in molecular chemistry~\cite{barron2004molecular}.
Recent advances in nanotechnology have enabled nanostructured systems exceeding the chiral responses of conventional materials~\cite{hildreth20123d,cui2014giant}. 
Moreover, planar chiral nanostructures have been developed to reduce the fabrication complexity and increase the capability to be integrated into optical circuits with other devices~\cite{du2015broadband,khanikaev2016experimental,lee2017microscopic}.
Yet, due to the inherent mirror symmetry of the 2-dimensional structures, achieving significant chiral signal in planar nanostructure at subwavelength scale is still challenging despite some successful reports. 

One of the promising approaches to the strong optical chirality of a planar nanodevice is using metallic nanostructures to support localized surface plasmon (LSP) resonances~\cite{ShuaiZu2016,hwang2017optical}.
LSP resonances are the oscillations of collective electrons which can be excited by light on the surfaces of metallic nanostructures.
The optical energy is confined to nanoscale volumes by the LSP resonances with potential for high-density integration of optical components~\cite{ozbay2006plasmonics,gramotnev2010plasmonics}.
Various nanodevices are designed and experimentally demonstrated based on LSP resonances exploiting the strong confinement and size reduction~\cite{kang2011low, lloyd2017plasmonic, hwang2016optical}.
In particular, the LSPs couple evanescently among one another when the metal nanoparticles are placed in proximity, which alters the resonance properties enabling intriguing optical phenomena such as plasmon induced transparency~\cite{zhang2008plasmon, liu2009plasmonic, hokari2014comparison}, plasmonic edge states~\cite{gomez2017plasmonic}, and Fano resonances~\cite{Fano1961, miroshnichenko2010fano}.

Fano resonances which occur by an interaction of two resonances with a narrow and a broad bandwidth~\cite{Fano1961}, where destructive and constructive interferences are observed depending on the wavelength~\cite{miroshnichenko2010fano}.
The characteristic asymmetric line shape of Fano resonances has also been found in the plasmonic particle systems supporting LSP resonances~\cite{lukyanchuk2010fano,rahmani2013fano}.
For instance, plasmonic Fano resonances appear in structures such as particle oligomers~\cite{mirin2009fano}, disk/ring assemblies~\cite{hao2008symmetry}, and nanoantenna assemblies~\cite{lovera2013mechanisms}.
Since their narrow spectral width and large field enhancement, plasmonic systems supporting Fano resonances have various applications including plasmonic rulers~\cite{liu2011three}, color routing~\cite{yan2017fano}, and biosensors~\cite{wu2012fano}.

Here, we present rigorous study on the effects of Fano resonance on the optical chirality in 2-dimensional plasmonic nanostructures both in experiment and theory.
There have been a theoretical study on circular dichroism induced by Fano resonances in planar chiral oligomers~\cite{Hopkins2016} and experimental demonstration on metallic arrays in the reflection optics configuration~\cite{ShuaiZu2016}.
Systematic analysis of Fano resonance of a single plasmonic chiral device, however, has not performed yet in the transmission optics configuration.
In our previous work~\cite{hwang2017optical}, we showed that the hexamers composed of gold nanorods placed on a glass substrate exhibit strong chiral response under dark-field illumination.
In this Letter, we systematically demonstrate the effects of Fano resonance on the optical chirality of subwavelength plasmonic nanodevices by introducing a nanodisk at the center of the hexamers.
Theoretical analysis using a coupled oscillator model and the numerical simulations are presented.
 
\begin{figure*}[htbp!]
  \includegraphics[width=0.82\textwidth]{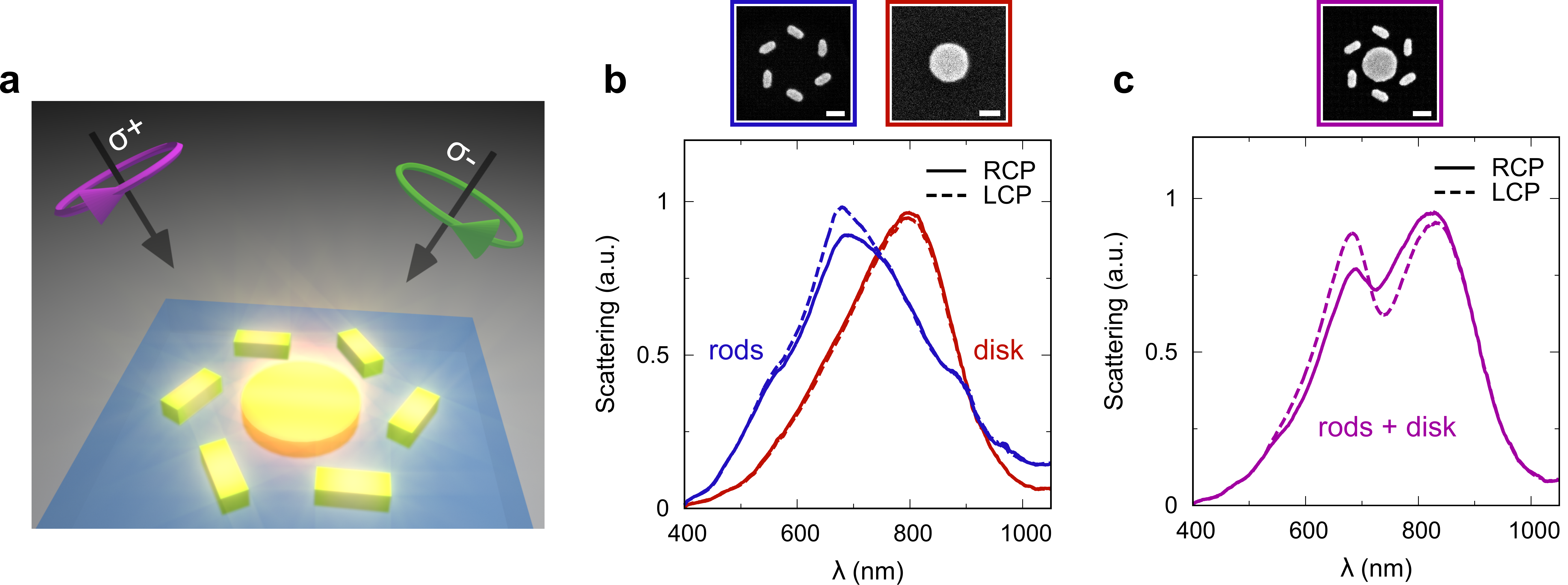}
  \caption{
    (a) A schematic of the dark-field illumination with two different spins onto the planar plasmonic nanostructure.
    (b) The scattering spectra of the hexamer of nanorods and the nanodisk from the dark-field illumination of differently rotating circularly polarized waves. The SEM images are displayed along with the spectra.
    (c) The scattering spectra of the combined nanodevice of the hexamer and the nanodisk along with the SEM image. The scale bar is 100 nm.
  }
  \label{rods_disk}
\end{figure*}

A combined structure of two plasmonic nanodevices with different resonance characteristics, one with and the other without optical chirality, is required to investigate the effects of Fano resonance on chiral signals so that the interference between the two resonance spectra can be controlled by varying the geometry of the achiral nanodevice.
Therefore, a plasmonic nanostructure consisting of six gold nanorods and one nanodisk is designed in order to investigate the effects of Fano resonance on chirality as shown in Fig.~\ref{rods_disk}(a).
As the hexamer shows strong chiral signal under dark-field illumination whereas the nanodisk shows the almost identical response for the spin states of the incident waves, the significance of Fano resonance can be engineered by changing the diameter of the nanodisk.
The schematic shown in Fig.~\ref{rods_disk}(a) artistically describes how the positive and negative spins of the incident light induce the LSP modes in the nanostructure.
The scattering spectra of the nanorods and the disk are separately measured under dark-field illumination and depicted in Fig.~\ref{rods_disk}(b).
The gold nanorod in the hexamer is 90~nm long, 34~nm wide, and 30~nm thick.
The diameter of the hexamer ring, the center-to-center distance between the two opposite nanorods, is 340~nm.
Each nanorod is rotated for $60^\circ$ to counter-clockwise from its radial axis.
The gold nanodisk has a diameter of 170~nm and a thickness of 30~nm.
The gold nanostructure adheres to a glass substrate by 2~nm germanium adhesion layer.
The hexamer of gold nanorods shows a peak at 680~nm in the scattering spectra for both RCP and LCP incident light due to the induced surface plasmons.
As a result of the existence of the glass substrate and the dark-field illumination, the plasmonic nanorods show the obvious difference between RCP and LCP in their scattering intensity as we reported in our previous research~\cite{hwang2017optical}.
On the other hand, the nanodisk presents a peak at 700~nm without chiral response as expected from its geometric symmetry.
The scattering spectra of the combined nanostructure of the rods and the disk are shown in Fig~\ref{rods_disk}(c).
A dip between the two peaks is clearly observed as a result of the coupling which is a characteristic of Fano resonance.
Remarkably, the change of the preference for the circular polarization is observed in the wavelength range of the disk resonance; scattering for RCP is greater than LCP in the range of $720-830$~nm while LCP scattering was stronger than RCP overall wavelength in the rods hexamer.

The plasmonic nanostructures were fabricated by means of electron-beam lithography (Vistec EBPG 5000 Plus ES) on glass substrates, using poly(methyl methacrylate) (Micro-Chem, 950k A2) and Copolymer (Micro-Chem, EL6) as a bilayer resist stack for a standard lift-off process and a sacrificial chromium layer (30~nm) that provides for charge dissipation during the exposure. 
The structures were developed with a 1:3 (by volume) mixture of methyl isobutyl ketone/2-propanol for 90~s, rinsed with 2-propanol, and dried with a nitrogen gun.
A 30~nm layer of Au was deposited by electron-beam evaporation, using 2 nm of germanium as the adhesion layer. 
A subsequent lift-off step with acetone produced the nanostructures.
Normal incidence images of the resulting structures were obtained by scanning electron microscopy (FEI, NovaNanoSEM 430).

The scattering spectra were acquired with a Nikon Ti-U microscope coupled to a cooled CCD camera (Princeton Instrument PIXIS).
The light source was a halogen lamp.
A linear polarizer and a quarter-wave plate and were placed before the dark-field condenser with the NA range of 0.80-0.95 in order to generate the circularly polarized dark-field incident wave.
The forward scattered light was collected using a 40x objective with NA of 0.60.

\begin{figure}[htbp!]
  \includegraphics[width=0.48\textwidth]{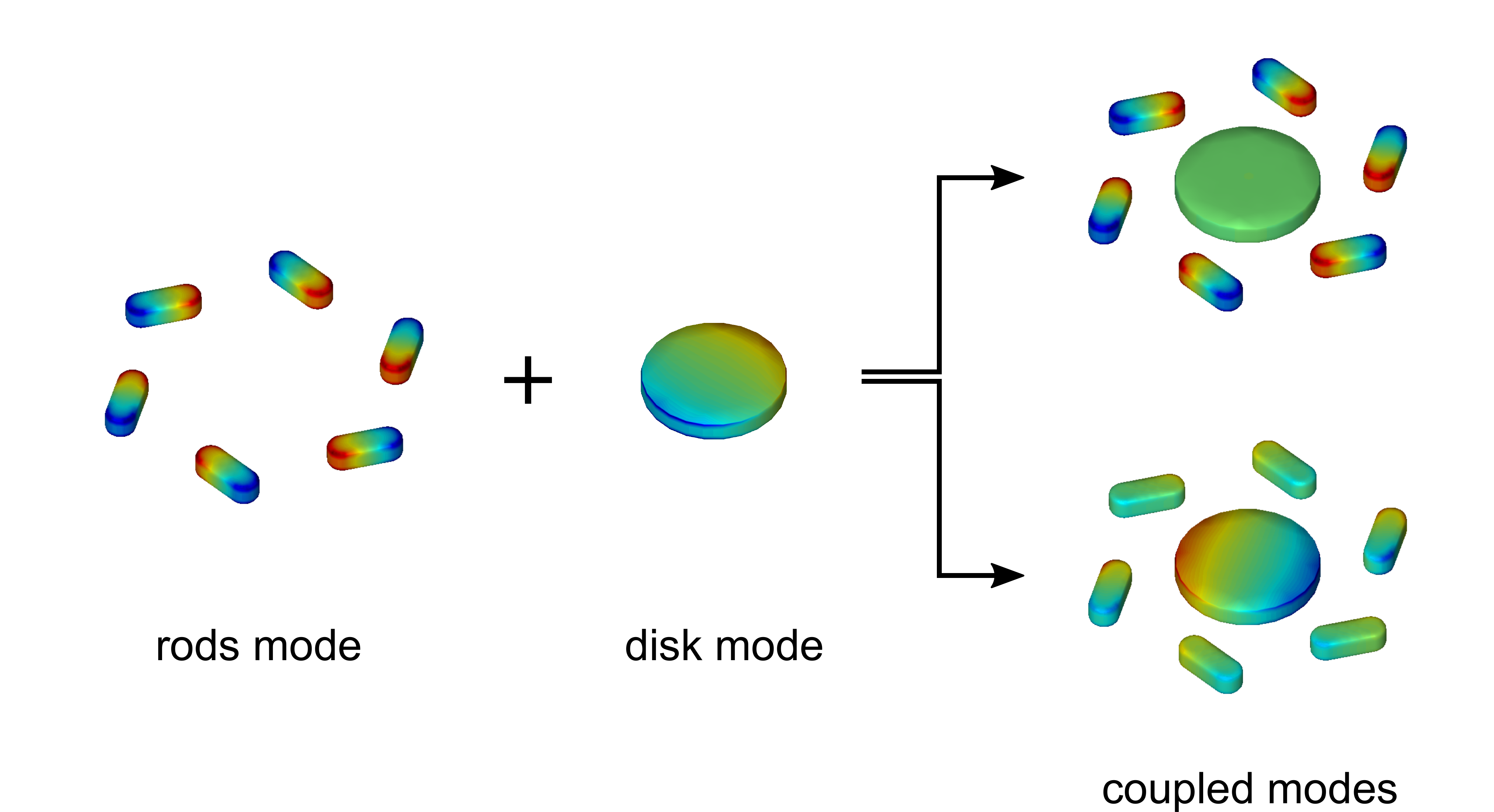}
  \caption{
    Surface charge distributions of the LSP modes supported by the rods, the disk, and the combined nanodevice obtained by the boundary element method.
  }
  \label{eem}
\end{figure}
We examined the mode characteristics of the nanodevice for a better understanding of the notable sign change of the chiral response.
Electrostatic eigenmode analysis is used to understand the characteristics of the modes supported by the combined nanostructure with the same physical dimensions of the experiment shown in Fig.~\ref{rods_disk}.
The dielectric function reported by Johnson and Christy~\cite{Johnson1972} is taken for gold.
First, the surface charge density profiles are obtained using the boundary element method~\cite{deabajo2002retarded,hohenester2012mnpbem} for the nanorod hexamer and the nanodisk separately (see Fig.~\ref{eem}).
Apparently, the LSP dipole mode of the individual nanorod is excited on each nanorod and their serial arrangement is the strongly supported mode which can be excited by a circularly polarized beam.
A dipole mode of LSP is also induced at the disk.
Next, the surface charge distributions of the combined structure are obtained.
Two representative coupled modes are generated by combining two structures which are rods dominant mode and a disk dominant mode shown on top and bottom, respectively.
Interestingly, the rods dominant coupled mode is not substantially affected by the nanodisk whereas the disk dominant coupled mode is affected by the presence of the nanorods.
It is expected that one of the two coupled modes will be more strongly excited depending on the spin of the incident light.

\begin{figure*}[hbtp!]
  \includegraphics[width=0.8\textwidth]{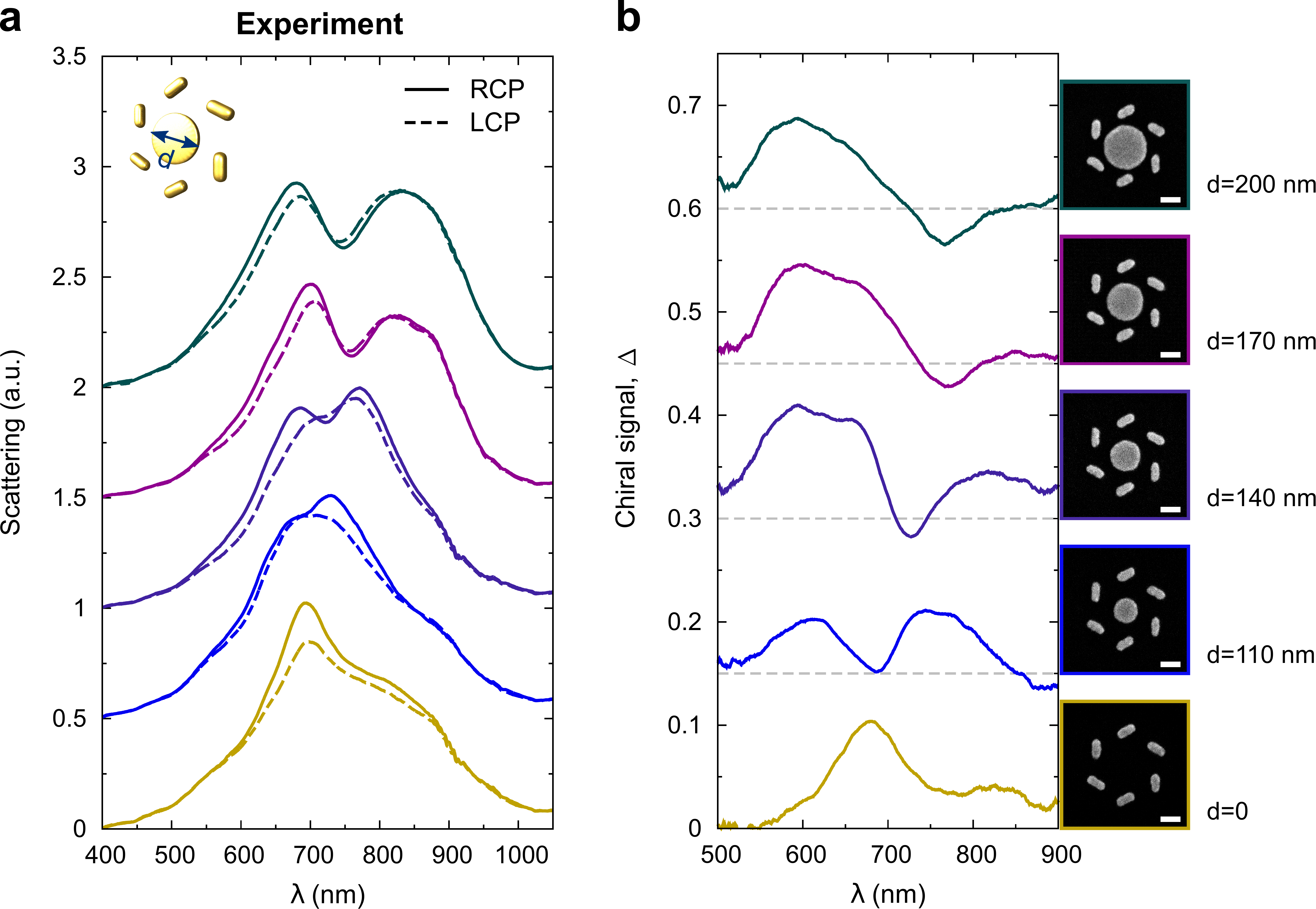}
  \caption{
    The measured (a) scattering and (b) chiral signal spectra for different disk diameters $d$.
    The SEM images of the fabricated gold nanostructures are displayed. 
    The scale bar is 100~nm.
    Gray dashed lines are shown in (b) to indicate the baselines of zero chiral signal of the corresponding nanodevices.
    The change to the opposite selection preference from RCP to LCP is clearly observed for the structures with $d=140, 170,$ and $200$~nm.
  }
  \label{disk_size}
\end{figure*}
We now demonstrate the effect of Fano resonance experimentally.
We introduced different size of nanodisks into the nanorod hexamer to examine the effect of Fano resonance by shifting the resonance wavelength of the disk as increasing the diameter.
The scattering spectra of the disk without the nanorods is shown in Supplementary Material.
The dimensions of the fabricated nanorods were measured to be $34\times90\,\mathrm{nm}^2$ while gold nanodisks were fabricated varying the diameter in the increment of $30$~nm, as shown in Figure~\ref{disk_size}b. 
The orientation angle of the nanorods is fixed to be $-60 ^\circ$.
The corresponding scattering and chirality spectra are provided in Figure~\ref{disk_size}a and b, respectively.
It is observed that the valley between two scattering peaks which is a characteristic feature of Fano resonance becomes more significant as the disk diameter increases.
When there is no disk at the center, the scattered intensity of RCP is greater than LCP over the measured wavelength range.
As Fano resonance becomes more significant by the increase of the disk size, however, the scattered intensity of LCP becomes greater than RCP in the longer wavelength.
To quantitatively analyze the chiral response, we define a chiral signal~\cite{hwang2017optical}
\begin{equation}
  \Delta = \frac{S_\mathrm{RCP} - S_\mathrm{LCP}}{S_\mathrm{RCP} + S_\mathrm{LCP}}, 
  \label{eq:Delta}
\end{equation}
where $S$ is the scattered intensity under dark-field illumination.
The experimentally obtained chiral signal spectra for different disk diameters are shown in Fig.~\ref{disk_size}(b).
The change to the opposite selection preference from RCP to LCP is clearly observed for the structures with $d=140, 170,$ and $200$~nm.
The observed negative $\Delta$ shifts to longer wavelength as increasing the disk diameter due to the red-shift of the disk mode.
 
\begin{figure*}[htbp!]
  \includegraphics[width=0.90\textwidth]{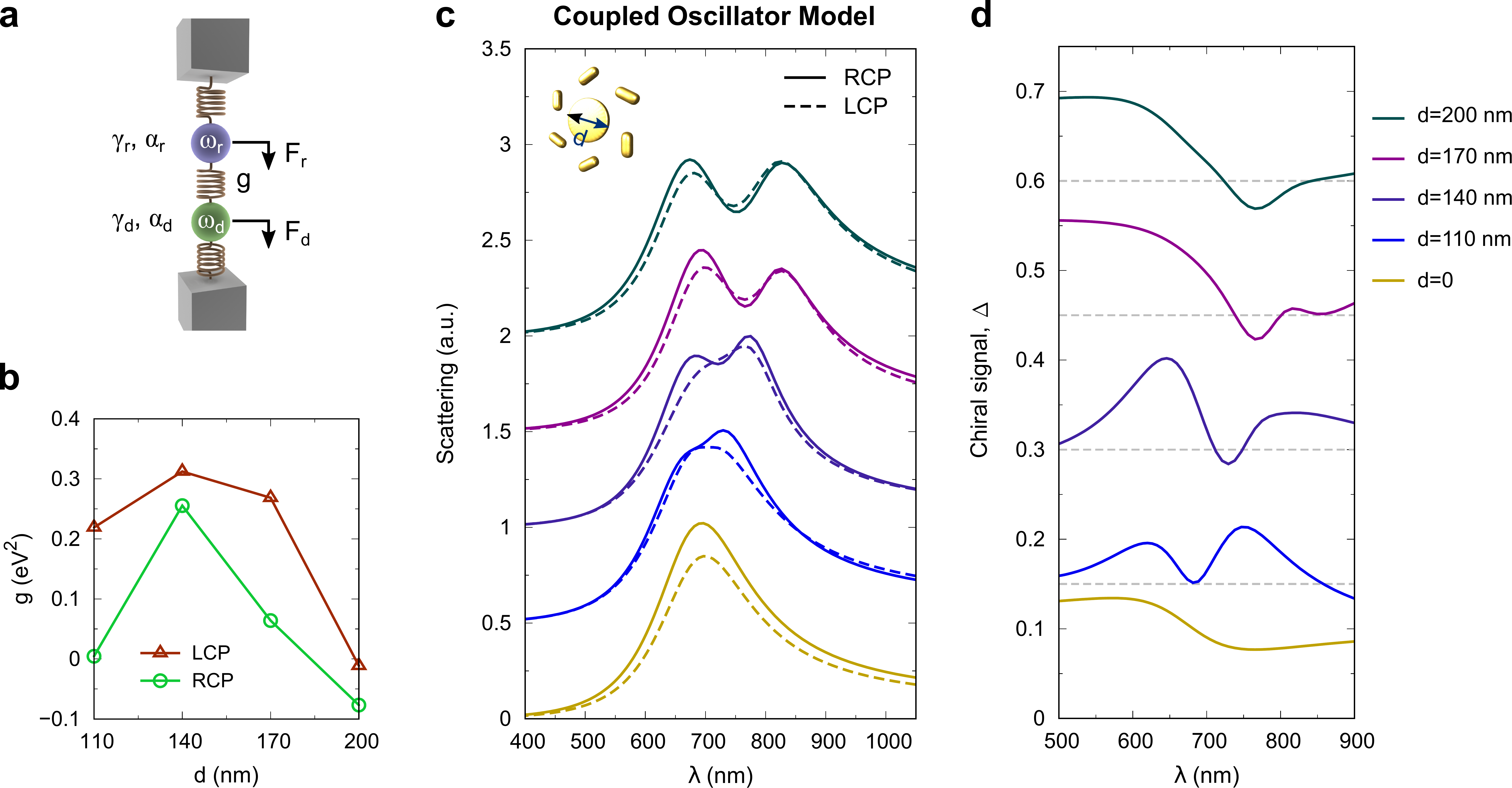}
  \caption{
    (a) Schematic of the extended coupled oscillator model.
    The top and the bottom oscillators represent the rods and the disk, respectively.
    (b) The coupling coefficient $g$ as a function of the disk diameter $d$ which has been obtained by fitting the model to the experimental results.
    The fit curves of the (c) scattering and (d) chiral signal spectra.
  }
  \label{model}
\end{figure*}
We present our theoretical analysis based on the extended coupled oscillator model in plasmonic systems studied by Lovera {\sl et al.}~\cite{lovera2013mechanisms}.
The schematic of the model is depicted in Fig.~\ref{model}(a).
The two oscillators in the schematic represent the rods and the disk with their resonant frequencies $\omega_r$ and $\omega_d$ and their damping coefficients $\gamma_r$ and $\gamma_d$, respectively.
The oscillators are coupled to each other with a coupling constant $g$.
The total dipole moment of the system is expressed as $P_{tot} = P_r + P_d = \alpha_r x_r + \alpha_d x_d$, where $P_{r,d}$, $x_{r,d}$, and $\alpha_{r,d}$ are the dipole moments, displacements, and the polarizabilities of the oscillators of the rods and the disk, respectively.
Consequently, the equations of motion are writtend as follows:
\begin{align}
\begin{aligned}
  \ddot{x}_r + \gamma_r \dot{x}_r + \omega_r^2 x_r + g x_d &= 0.5 \dddot{P}_{tot} + F_r \\
  \ddot{x}_d + \gamma_d \dot{x}_d + \omega_d^2 x_d + g x_r &= 0.5 \dddot{P}_{tot} + F_d,
\end{aligned}
\label{eq}
\end{align}
where $F_{r,d}=\alpha_{r,d} E_{ext}$ which is the net force applied on each oscillator.

We assume an incident harmonic field, $E_{ext} = E_0 e^{i\omega t}$, which induces consequent displacements of $x_{r,d}(\omega)=C_{r,d}(\omega) e^{i\omega t}$, where $C_{r,d}$ are the oscillation amplitudes.
Once $C_{r,d}$ are computed analytically from \eqref{eq}, the scattered intensity can be obtained by calculating the squared modulus of the total amplitude, $\lvert C_r + C_d \rvert^2$.
In Figs.~\ref{model}(c) and (d), the curves fit the experimental spectra using the coupled oscillator model are shown.
The characteristic dips of Fano resonance are clearly shown in the scattering spectra.
The coupling coefficient $g$ extracted from the model is shown as a function of the disk diameter $d$ in Fig.~\ref{model}(b).
It is observed that the obtained $g$ is higher at LCP than at RCP illumination over the measured disk diameter range, which indicates that the coupling of the two resonators shows a chiral response to the dark-field illumination with an apparent preference.
The positive and negative $g$ mean the in-phase and the out-of-phase oscillations of the two resonators, respectively.
It is worth to note that $g<0$ obtained when $d=200$~nm for both circular polarizations can be understood as the dipole modes induced at the disk oscillate out-of-phase to the LSP modes at the rods.
The entire list of the extracted parameters is shown in Table~\ref{table1}.
The resonant frequencies of the disk $\omega_d$ show a red-shift as the disk size increase, while the change of $\omega_r$ is insignificant. 
The selection preference transition from $\Delta>0$ to $\Delta<0$ can be understood as a combined effect of this red-shift and the coupling of the two oscillators.

\begin{table*}
  \centering
  \caption{Extracted parameters from the coupled oscillator model for varying disk diameters (Fig.~\ref{model})}
  \begin{tabular}{cccccccccc}
    \\
    $d$(nm) & pol & $\omega_r$(eV) & $\omega_d$(eV) & $g$(eV$^2$) & $E_0$
    & $\gamma_r$(eV) & $\gamma_d$(eV) & $\alpha_r$(eV$^{-1}$) & $\alpha_d$(eV$^{-1}$) \\[1mm] \hline 
    \multirow{2}{*}{  0} & RCP & 1.842 & -     & -     & 12.09 & 0.392 & -     & 0.079 & -     \\ 
                         & LCP & 1.827 & -     & -     & 10.82 & 0.385 & -     & 0.079 & -     \\ \hline
    \multirow{2}{*}{110} & RCP & 1.862 & 1.732 &~0.045 & 14.47 & 0.375 & 0.256 & 0.051 & 0.015 \\ 
                         & LCP & 1.840 & 1.797 &~0.219 & ~4.03 & 0.150 & 0.157 & 0.126 & 0.140 \\ \hline
    \multirow{2}{*}{140} & RCP & 1.829 & 1.644 &~0.256 & 11.61 & 0.348 & 0.191 & 0.036 & 0.036 \\ 
                         & LCP & 1.802 & 1.642 &~0.312 & ~9.61 & 0.438 & 0.177 & 0.049 & 0.398 \\ \hline 
    \multirow{2}{*}{170} & RCP & 1.800 & 1.544 &~0.064 & 12.78 & 0.334 & 0.212 & 0.052 & 0.017 \\ 
                         & LCP & 1.774 & 1.550 &~0.269 & 12.47 & 0.345 & 0.221 & 0.032 & 0.321 \\ \hline 
    \multirow{2}{*}{200} & RCP & 1.846 & 1.547 &-0.077 & 28.41 & 0.414 & 0.279 & 0.030 & 0.006 \\ 
                         & LCP & 1.837 & 1.543 &-0.011 & 26.85 & 0.396 & 0.298 & 0.025 & 0.010 \\ \hline
  \end{tabular}
  \label{table1}
\end{table*}

\begin{figure*}[htbp!]
  \includegraphics[width=0.7\textwidth]{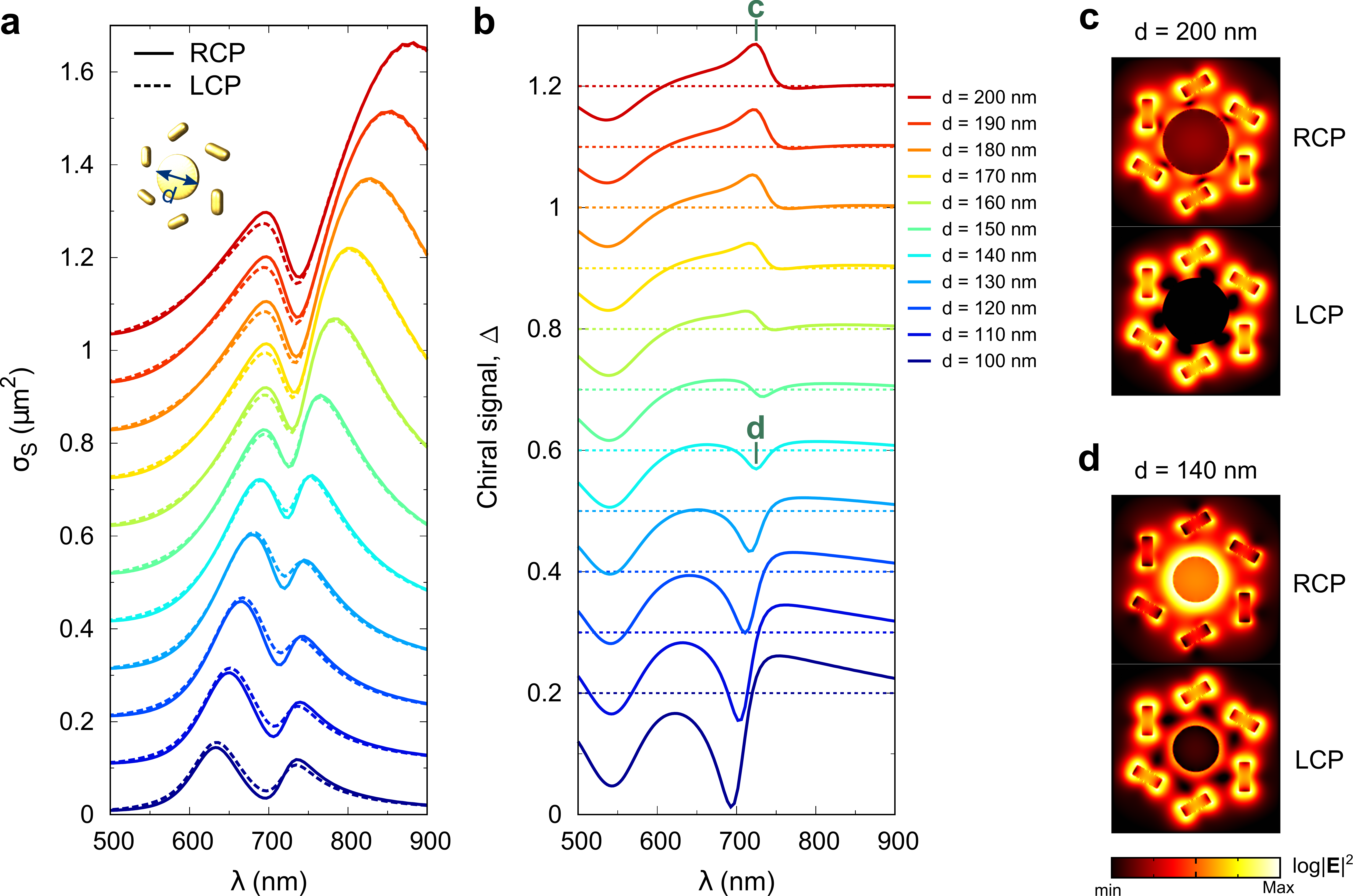}
  \caption{
    FDTD simulation results of (a) scattering cross-section and (b) chiral signal spectra for different disk diameters $d$.
    Color dashed lines are shown in (b) to indicate the baselines of zero chiral signal of the corresponding nanodevices.
    The electric-field intensity profiles of the nanostructures with (c) $d=200$~nm and (d) $d=140$~nm at the wavelength $\lambda=725$~nm as indicated in (b).
}
  \label{fdtd}
\end{figure*}

Three-dimensional full-wave simulations using the finite-difference time-domain (FDTD) method (Lumerical Inc.) have been performed for the further understanding of the effects of Fano resonance caused by the nanodisk at the center of the nanostructure, and the obtained scattering and the chiral signal spectra are shown in Fig.~\ref{fdtd}.
The complex refractive index reported by Johnson and Christy is used to determine the optical parameters of the plasmonic nanostructures~\cite{Johnson1972}. 
The dark-field illumination with circular polarization is created using four linearly polarized Gaussian sources. 
The circularly polarized light is generated by interfering two linearly polarized waves with $\pi/2$ phase difference. 
Afterward, two beams with different numerical apertures (NA), 0.95 and 0.80,  are employed to create a dark-field beam for each circular polarization. 
The two beams are in antiphase ($\pi$ phase difference) for the beam of smaller NA to be subtracted from that of larger NA to form a ring-shaped incident beam in consequence.
The total-field scattered-field method is applied to obtain the scattered intensity by the nanodevices.
The diameter of the central disk is varied in the range of $d=100-200$~nm with the increment of 10~nm.
The resonance wavelength of the hexamer without the central disk obtained by the simulation is 725~nm both for RCP and LCP (see Supplementary Material).
The scattering cross-section of the RCP is obtained greater than the LCP in the wavelength range of $680-900$~nm which agrees with the experimental results.
On the other hand, the calculated resonance wavelength of the nanodisk with the nanorods shifts from $635-855$~nm as increasing the disk diameter.
The two resonances caused by the nanorods and the nanodisk overlap the most when the disk diameter $d$ is 140~nm with the resonance wavelength of 715~nm.
In result, the scattering spectra at $d=140$~nm show two peaks with similar scattering cross-section.
Interestingly, the corresponding chiral signal spectrum shows a dip at the resonance wavelength which indicates the reversed preference of rotational direction of the circularly polarized light.
The electric field intensity of the given structure at $\lambda=725$~nm is depicted in Fig.~\ref{fdtd}(d), where the obvious difference between the RCP and the LCP is shown: The disk mode is dominant in RCP while the rods mode in LCP.
It should be noted that the disk mode itself has no preference for RCP and LCP, thus the preference originated from the coupling with the rods mode.
On the contrary, the resonance of the disk with $d=200$~nm has little overlap with the rods mode, the rods mode is dominant both for RCP and LCP as displayed in Fig.~\ref{fdtd}(c).
Subsequently, the chiral signal of the nanodevice at the given wavelength has the same positive sign as the hexamer without the disk.

To conclude, we have observed an intriguing transition to the opposite selection preference of different handedness of the circularly polarized light which is originated from the Fano resonance.
We have systematically demonstrated the effects of Fano resonance on the chiroptical response of subwavelength plasmonic nanodevices.
Theoretical analysis using a coupled oscillator model has explained the effects of Fano resonance in terms of the coupling between the rods and the disk modes.
FDTD simulations supported the experimental results and the theoretical model.
The dependence of the diameter of the nanodisk has been studied both theoretically and experimentally.
This study enriches the scientific understanding of the effects of Fano resonance in two-dimensional plasmonic chiral systems and provides insights on optical chirality in coupled systems including destructive interferences.

\begin{acknowledgments}
This work was partially supported by the National Natural Science Foundation of China under Grant Nos. 61490712, 61427819, U1701661; National Key Basic Research Program of China (973) under grant No.2015CB352004; the leading talents of Guangdong province program No. 00201505; the Natural Science Foundation of Guangdong Province under No.2016A030312010; and the Science and Technology Innovation Commission of Shenzhen under grant Nos. KQTD2015071016560101, ZDSYS201703031605029.
This work was performed in part at the Melbourne Centre for Nanofabrication (MCN) in the Victorian Node of the Australian National Fabrication Facility (ANFF).
YH acknowledges the discussions on theory with B. Hopkins.
\end{acknowledgments}

\bibliography{fano}

\end{document}


\title{Supplemental Material for ``Effects of Fano resonance on optical chirality of planar plasmonic nanodevices''}

\author{Yongsop Hwang}
\affiliation{Nanophotonics Research Centre, Shenzhen University, Shenzhen, 518060, China}
\affiliation{School of Engineering, RMIT University, Melbourne, VIC 3001, Australia}
\author{Seojoo Lee}
\affiliation{Department of Physics, Korea University, Seoul 02841, Republic of Korea}
\author{Sejeong Kim}
\affiliation{School of Mathematical and Physical Sciences, University of Technology Sydney, Ultimo, NSW 2007, Australia}
\author{Q-Han Park}
\affiliation{Department of Physics, Korea University, Seoul 02841, Republic of Korea}
\author{Jiao Lin}
\email{jiao.lin@rmit.edu.au}
\affiliation{Nanophotonics Research Centre, Shenzhen University, Shenzhen, 518060, China}
\affiliation{School of Engineering, RMIT University, Melbourne, VIC 3001, Australia}
\author{Xiao-Cong Yuan}
\email{xcyuan@szu.edu.cn}
\affiliation{Nanophotonics Research Centre, Shenzhen University, Shenzhen, 518060, China}

\date{\today}
\maketitle

In this Supplementary Material, the investigation of the rods and the disk as separate systems are presented. Experimental and theoretical studies are included.

\section{Nanodisk without the rods}

\begin{figure*}[htbp!]
  \includegraphics[width=0.9\textwidth]{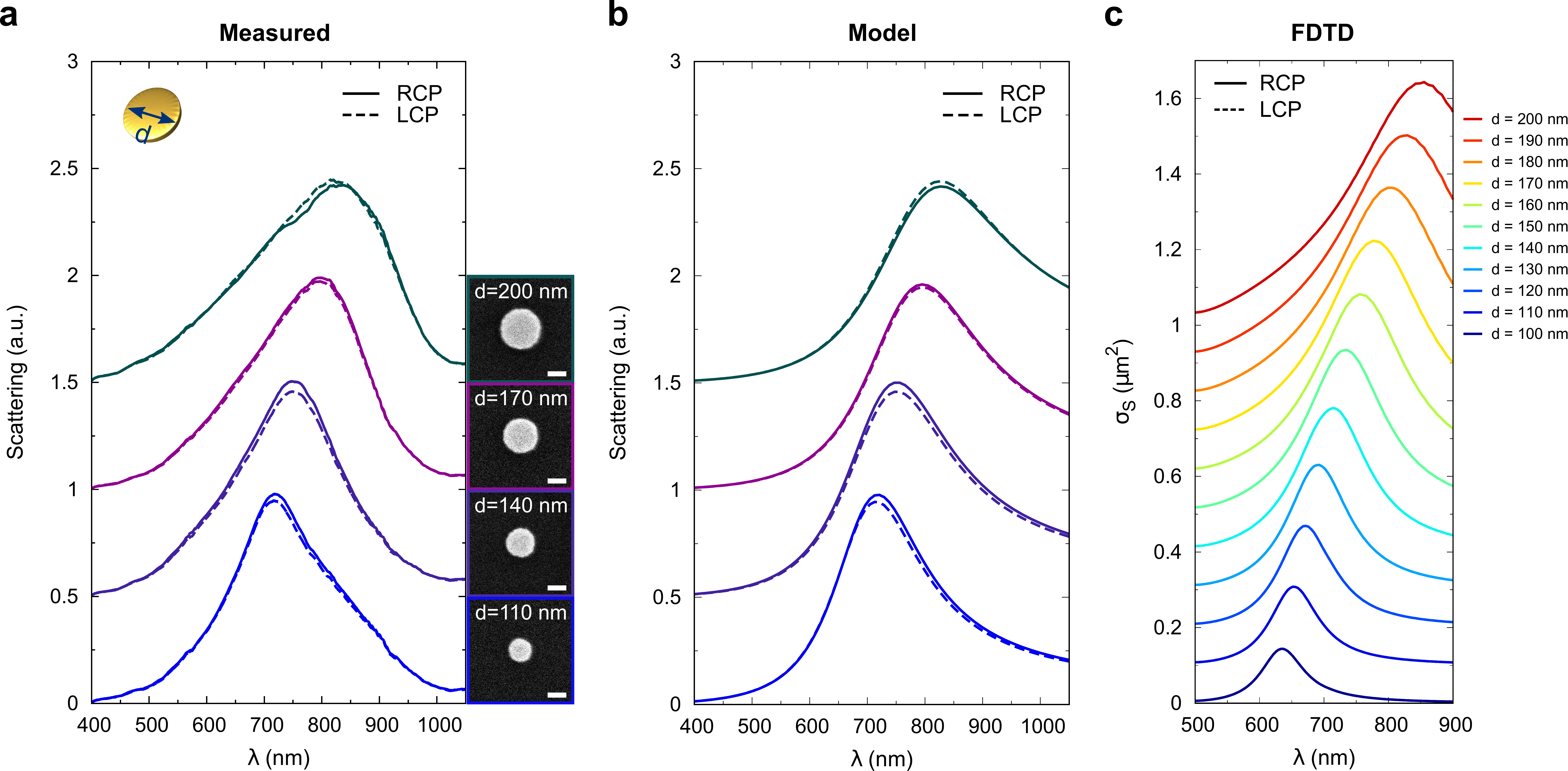}
  \caption{
  The red-shift of the plasmonic resonance of the nanodisk by increasing its diameter.
  (a) Experiment, (b) oscillator model, and (c) FDTD simulation.
  The scale bar in the SEM images is 100~nm.
  }
  \label{diskonly}
\end{figure*}

The gold nanodisk without the nanorod hexamer is studied experimentally by varying its diameter.
The plasmonic resonance wavelength of the dark-field scattering measured to shift to the longer wavelength as the disk size increases as shown in Fig.~\ref{diskonly}(a).
The theoretical modeling to fit the measured spectra is performed as shown in Fig.~\ref{diskonly}(b) and the corresponding parameters are listed in Table~\ref{tab:diskonly}.
The FDTD simulation results also show the red-shift agreeing with the experiment.
The provided resonance characteristics of the nanodisk are expected to assist the understanding of the Fano resonances in the main text.

\begin{table*}
  \centering
  \caption{Extracted parameters from the oscillator model for varying disk diameters of the nanodisk without the rods (Fig.~\ref{diskonly})}
  \begin{tabular}{cccccc}
    \\
    $d$(nm) & pol & $\omega_d$(eV) & $E_0$ & $\gamma_d$(eV) & $\alpha_d$(eV$^{-1}$) \\[1mm] \hline 
    \multirow{2}{*}{110} & RCP & 1.774 & 11.01 & 0.364 & 0.075 \\ 
                         & LCP & 1.781 & 11.07 & 0.369 & 0.075 \\ \hline
    \multirow{2}{*}{140} & RCP & 1.704 & 13.21 & 0.428 & 0.066 \\ 
                         & LCP & 1.704 & 12.94 & 0.428 & 0.066 \\ \hline 
    \multirow{2}{*}{170} & RCP & 1.612 & 12.48 & 0.431 & 0.064 \\ 
                         & LCP & 1.611 & 12.45 & 0.431 & 0.063 \\ \hline 
    \multirow{2}{*}{200} & RCP & 1.557 & 14.03 & 0.481 & 0.057 \\ 
                         & LCP & 1.559 & 14.02 & 0.472 & 0.057 \\ \hline
  \end{tabular}
  \label{tab:diskonly}
\end{table*}

\section{Nanorods without the disk}

\begin{figure*}[tbp!]
  \includegraphics[width=0.8\textwidth]{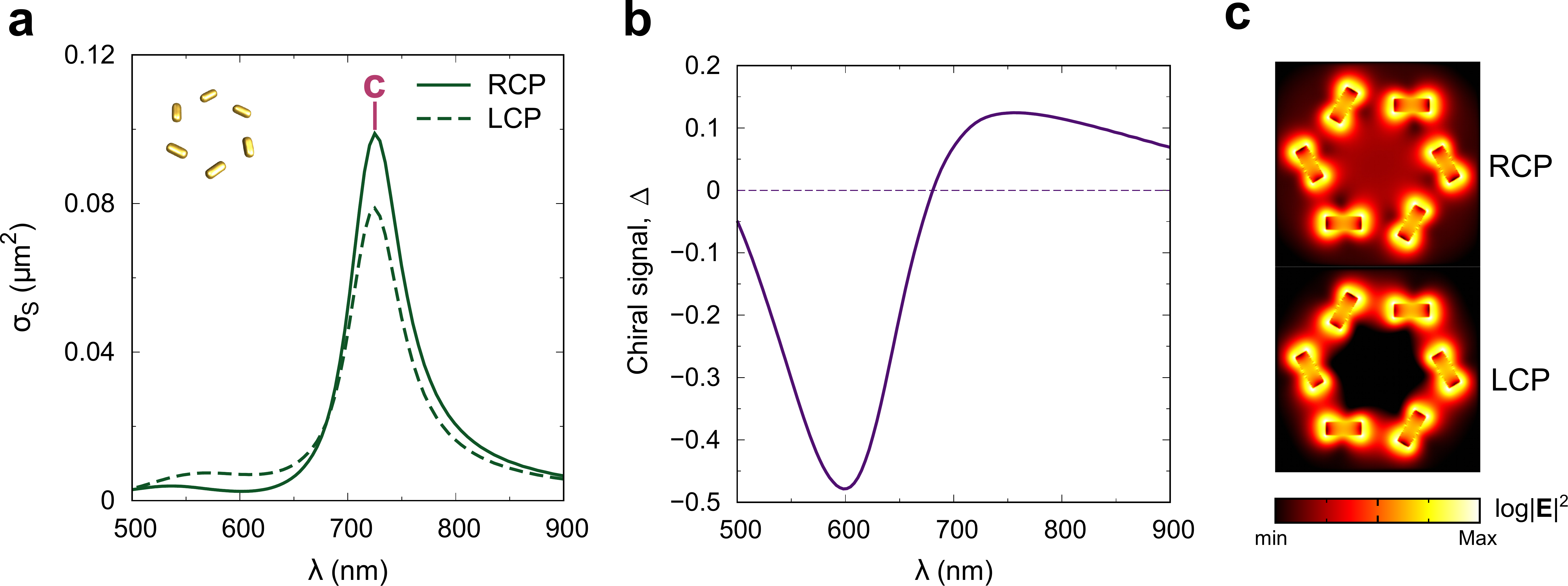}
  \caption{
  FDTD simulation of the hexamer of nanorods.
  (a) Scattering spectra of RCP and LCP.
  (b) Chiral signal.
  (c) E-field intensity profile in log scale.
  }
  \label{fdtd_rods}
\end{figure*}

The FDTD simulation results of the nanorod hexamer without the disk is provided in Fig.~\ref{fdtd_rods}.
The experimental data of the hexamer is presented in the main text as $d=0$ along with those with the other disk diameters.
The chirality is clearly observed in the spectra and the field profiles at the resonance peak $\lambda=725$~nm are displayed in Fig.~\ref{fdtd_rods}(c).

